\documentclass[iop]{emulateapj}
\usepackage{apjfonts}

\newcommand{\Pe}{P_{\mathrm{e}}}
\newcommand{\Pc}{P_{\mathrm{c}}}
\newcommand{\nel}{n_{\mathrm{e}}}
\newcommand{\Pec}{P_{\mathrm{ec}}}
\newcommand{\Binit}{B_{\mathrm{i}}}
\newcommand{\Bfin}{B_{\mathrm{f}}}
\newcommand{\Bcrit}{B_{\mathrm{crit}}}
\newcommand{\EFec}{E_{\mathrm{F, ec}}}
\newcommand{\Sigmaecinit}{\Sigma_{\mathrm{ec, i}}}
\newcommand{\Sigmaecfin}{\Sigma_{\mathrm{ec, f}}}
\newcommand{\mue}{\mu_{\mathrm{e}}}
\newcommand{\hec}{H_{\mathrm{ec}}}
\newcommand{\Etot}{E_{\mathrm{ec}}}
\newcommand{\Emag}{E_{B}}
\newcommand{\Eper}{E_{\mathrm{int}}}
\newcommand{\pperp}{p_{\perp}}
\newcommand{\tage}{\tau_{\mathrm{char}}}

\renewcommand{\vec}[1]{\mbox{$\mathbf{#1}$}}
\newcommand{\grad}{\vec{\nabla}}
\newcommand{\vdot}{\vec{\cdot}}

\newcommand{\ee}[1]{\ensuremath{\times 10^{#1}}}

\newcommand{\EF}{\ensuremath{E_\mathrm{F}}} 
\newcommand{\mb}{\ensuremath{m_\mathrm{u}}}
\newcommand{\me}{\ensuremath{m_\mathrm{e}}} 

\newcommand{\unitspace}{\ensuremath{\,}}
\newcommand{\usp}{\unitspace}
\newcommand{\numberspace}{\ensuremath{\;}}
\newcommand{\nsp}{\numberspace}
\newcommand{\unitstyle}[1]{\ensuremath{\mathrm{#1}}}
\newcommand{\power}[2]{\ensuremath{{#1}^{#2}}}

\newcommand{\centi}{\unitstyle{c}}
\newcommand{\kilo}{\unitstyle{k}}
\newcommand{\Mega}{\unitstyle{M}}

\newcommand{\meter}{\unitstyle{m}}

\newcommand{\second}{\unitstyle{s}}

\newcommand{\cm}{\centi\meter}

\newcommand{\dyne}{\unitstyle{dyne}}
\newcommand{\erg}{\unitstyle{erg}} 
\newcommand{\ergs}{\erg}
\newcommand{\gauss}{\unitstyle{G}}
\newcommand{\ergspersecond}{\erg\unitspace\power{\second}{-1}}
\newcommand{\eV}{\unitstyle{eV}}
\newcommand{\keV}{\kilo\eV} 
\newcommand{\MeV}{\Mega\eV} 
\newcommand{\Msun}{\ensuremath{M_\odot}}
\newcommand{\yr}{\unitstyle{yr}}      
\newcommand{\km}{\kilo\meter}

\begin{document}
\title{Magnetic Field-Decay-Induced Electron Captures: a Strong Heat Source in Magnetar Crusts}
\author{Randall L. Cooper and David L. Kaplan\altaffilmark{1}}
\affil{Kavli Institute for Theoretical Physics, University of California, Santa Barbara, CA 93106, USA; rcooper@kitp.ucsb.edu, dkaplan@kitp.ucsb.edu}
\altaffiltext{1}{Hubble Fellow}

\begin{abstract}

We propose a new heating mechanism in magnetar crusts.  Magnetars' crustal magnetic fields are much stronger than their surface fields; therefore, magnetic pressure partially supports the crust against gravity.  The crust loses magnetic pressure support as the field decays and must compensate by increasing the electron degeneracy pressure; the accompanying increase in the electron Fermi energy induces nonequilibrium, exothermic electron captures.  The total heat released via field-decay electron captures is comparable to the total magnetic energy in the crust.  Thus, field-decay electron captures are an important, if not the primary, mechanism powering magnetars' soft X-ray emission.

\end{abstract}

\keywords{nuclear reactions, magnetic fields --- stars: neutron}

\section{Introduction}\label{s.introduction}

Soft gamma repeaters (SGRs) and anomalous X-ray pulsars \citep[AXPs; for reviews, see][]{WT06,K07,M08} are the two known manifestations of magnetars: neutron stars with ultrastrong exterior magnetic fields $B \gtrsim 10^{14} \nsp \gauss$ and even stronger interior, primarily toroidal fields \citep{DT92,P92,TD93}.  As of this writing, all magnetars have characteristic ages $\tage \lesssim 10^{5} \nsp \yr$, and their persistent soft X-ray luminosities $L \sim 10^{35} \nsp \ergspersecond$ \citep[although some are transients, e.g.,][]{Ietal04}, which are likely of thermal origin, are systematically higher by factors of $>100$ than the power available from rotation \citep{vPTvdH95}, cooling \citep[e.g.,][] {PGW06}, or accretion \citep{HvKK00,Ketal01}.  Taken together, these observations suggest that magnetars' magnetic fields decay and thereby induce heating in the stellar interiors \citep[e.g.,][]{TD96,HK98,CGP00,PG07,PLMG07}.

Sustaining the persistent thermal emission requires an interior energy reservoir 
\begin{equation}\label{e.persistentE}
\Eper \gtrsim 3 \ee{47} \left ( \frac{\tage}{10^{5} \nsp \yr} \right ) \left (  \frac{L}{10^{35} \nsp \ergspersecond} \right ) \nsp \ergs.
\end{equation}
Comparing this to the magnetar's total internal magnetic energy $\Emag \sim (B^{2}/8 \pi) (4 \pi R^{3}/3)$ gives $B \gtrsim 1\ee{15} \nsp \gauss$.  However, these values are likely flagrant underestimations:  First, a magnetar emits most of its thermal energy as neutrinos.  Indeed, \citet{KYPSSG06,KPYC09} argue that, to avoid severe neutrino losses, the heat source powering the thermal emission must be located at or near the outer crust, i.e., within the magnetar's outermost $\approx 100 \nsp \meter$.  If the thermal emission originates from magnetic energy contained only within the outer crust, then $B \sim 10^{16} \nsp \gauss$.  Second, a magnetar's magnetic field powers other phenomena as well, such as giant flares;  the 2004 December 27 giant flare from SGR 1806--20 released $\sim 10^{46} \nsp \ergs$, which implies $B \sim 10^{16} \nsp \gauss$ \citep[assuming giant flares are recurrent;][]{Hetal05,SDIV05,Tetal05}.  

That the magnetic field's strength is much greater in the crust than at the surface implies that a substantial $B$ gradient exists.  Therefore, if the crustal field geometry is primarily nonradial, magnetic pressure partially supports the crust against gravity.  We argue in \S \ref{s.fielddecayecaptures} that the crust loses magnetic pressure support as $B$ decays and compensates with an increase in electron degeneracy pressure.  The accompanying increase in the electron Fermi energy induces exothermic electron captures that heat the crust.  We calculate the total energy released via field-decay electron captures in \S \ref{s.crustalheating} and compare it to Joule heating in \S \ref{s.jouleheatingcomparison}.  We discuss our results and speculate on other consequences of field-decay electron captures in \S \ref{s.discussion}.

\section{Field-Decay Electron Captures}\label{s.fielddecayecaptures}

We construct a simple model of the hydrostatic structure of a neutron star's strongly magnetized outer crust to illustrate the process by which magnetic field decay induces electron captures.  We ignore the inner crust both for simplicity and because most of the thermal energy released there escapes as neutrinos.  Consider a neutron star of mass $M$ and radius $R$.  The outer crust's scale height \citep[$\approx 100 \nsp \meter$, e.g.,][]{ST83} is much less than $R$; therefore, we adopt a plane-parallel geometry and set the gravitational acceleration $g = GM/R^{2}$ to be constant throughout the layer.  Furthermore, we make three assumptions regarding the crustal magnetic field: (1) Its nonradial (e.g., toroidal) component is large relative to its radial component, (2) the field strength in the crust is greater than that at the stellar surface, and (3) the field decays with time.  These assumptions are likely satisfied in magnetars \citep{TD93,TD95,TD96,TD01,BS06,GKP06,PMP06,PPMM06,B09,R09}.

The equation of hydrostatic equilibrium in the electron magnetohydrodynamic limit is
\begin{equation}\label{e.hydroeq}
\rho \vec{g} - \grad P  - \grad \left (\frac{B^{2}}{8\pi} \right ) + \frac{1}{4\pi} (\vec{B} \, \vdot \grad )\vec{B} = \vec{0},
\end{equation}
where $\rho$ is the density, $P$ is the pressure, and $\vec{B}$ is the magnetic field.  We ignore the magnetic field's affect on $P$ (see Appendix).  Setting the radial component of $\vec{B}$ equal to zero by assumption (1) for simplicity, the equation of vertical hydrostatic equilibrium becomes
\begin{equation}\label{e.verthydroeq}
\rho g = - \frac{\partial}{\partial r} \left ( P + \frac{B^{2}}{8\pi} \right ).
\end{equation}
Invoking the plane-parallel geometry and rewriting Equation (\ref{e.verthydroeq}) in terms of the Lagrangian coordinate $\Sigma$, the column depth as measured from the surface (such that $d \Sigma = - \rho dr$),
\begin{equation}\label{e.verthydroeq2}
g =  \frac{\partial}{\partial \Sigma} \left ( P + \frac{B^{2}}{8\pi} \right ).
\end{equation}
Finally, integrating Equation (\ref{e.verthydroeq2}) from the surface, we get
\begin{equation}\label{e.verthydroeq3}
g \Sigma =  P (\Sigma, t) + \frac{1}{8\pi} [B^{2} (\Sigma,t) - B^{2} (0,t)],
\end{equation}
where we have set $P = 0$ at the surface, where $\Sigma = 0$, and included the explicit time and spatial dependences of $P$ and $B$.  Equation (\ref{e.verthydroeq3}) shows that electron degeneracy pressure and the magnetic field together support the column of matter $\Sigma$ against gravity.  

Consider a matter element at column depth $\Sigma$ containing a nucleus of mass $M(A,Z)$, where $A$ is the mass number and $Z$ is the proton number, as $B$ decays with time.  The column gradually loses magnetic pressure support; to compensate, $P$, and thereby the electron Fermi energy $\EF$, increases with time.  When $M(A,Z)c^{2} + \EF > M(A,Z-1)c^{2}$, electron capture onto the nucleus becomes energetically favorable.  

The nuclear species at a given $\Sigma$ is a function of $\EF$ \citep[or equivalently, $P$; e.g.,][]{ST83}, so even when the crust is initially at equilibrium, electron captures will occur: Any nucleus that is stable at a given $\EF$ is unstable to capture at a sufficiently larger $\EF$. The loss of magnetic pressure support due to field decay forces nuclei out of their initial equilibrium and into a new equilibrium configuration.

Electron captures can be exothermic in two mutually inclusive ways \citep[e.g.,][]{GBSMK07}: (1) An electron captures into an exited state of the daughter nucleus if, for example, the daughter nucleus's ground state is forbidden.  The daughter nucleus then radiatively deexcites and thereby heats the crust.  (2) If the parent nucleus is even-even, then $M(A,Z-1) > M(A,Z-2)$ due to the nuclear pairing energy, and a second, nonequilibrium electron capture immediately ensues and releases heat.  

This process resembles the well-known deep crustal heating in accreting neutron stars \citep[e.g.,][]{BBR98,GBSMK07,GKM08,HZ08}:  In that context, $P$ alone supports the matter column against gravity, so $g \Sigma = P$.  The accretion-driven advection of a matter element to higher pressures, and thereby higher $\EF$, induces electron captures that heat the crust.  In magnetars, however, it is not accretion, but magnetic field decay, that induces electron captures\footnote{The effective gravitational acceleration increase during spin-down of a rapidly rotating neutron star raises $\EF$ of a matter element and thus induces electron captures as well, although it is unimportant for SGRs and AXPs because they rotate slowly.  This effect is similar to rotochemical heating in neutron star cores \citep{R95}.}.

\section{Crustal Heating Via Field-Decay Electron Captures}\label{s.crustalheating}

We now estimate the total thermal energy released in the outer crust via field-decay electron captures.  Consider a neutron star crust with an initial magnetic field $\Binit$ and that contains an ion species with an electron capture threshold energy $\EFec$, such that electrons capture when $\EF > \EFec$.  After a period of time, the crustal magnetic field decays (via unspecified means) to $\Bfin \ll \Binit$.  For relativistic, degenerate electrons, $P \propto \EF^{4}$; we define $\Pec$ to be the pressure at which $\EF = \EFec$.

\begin{figure}
\epsscale{1.1}
\plotone{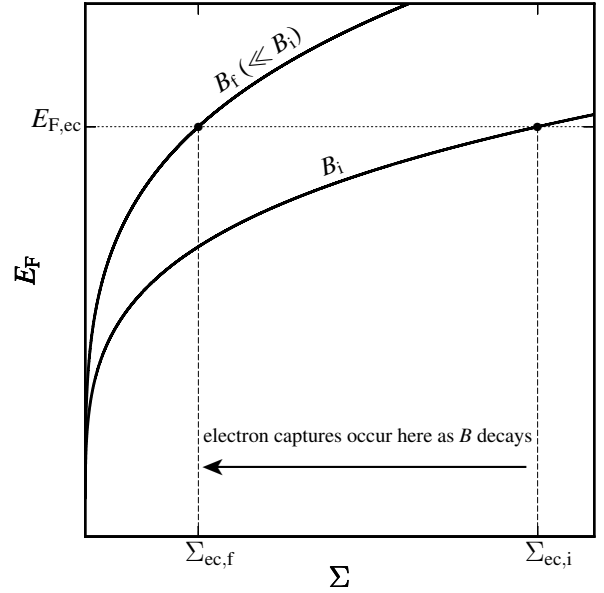}
\caption{Schematic diagram of the electron Fermi energy $\EF$ as a function of column depth $\Sigma$ during crustal magnetic field decay.  Electron degeneracy pressure $P$ and magnetic pressure $B^{2}/8\pi$ together support the column against gravity.  The column loses magnetic pressure support as $B$ decays from $\Binit$ to $\Bfin$, and $P$ and $\EF$ concomitantly increase.  When $\EF > \EFec$, the electron capture threshold, exothermic captures ensue and heat the crust.  The matter column within which field-decay electron captures occur is $\Sigmaecinit - \Sigmaecfin$.
}
\label{f.schematic}
\end{figure}

We now determine the column of matter within which electrons capture as the field decays from $\Binit$ to $\Bfin$.  The vertical hydrostatic equilibrium equations for columns of matter at the electron capture pressure $\Pec$ before and after field decay are (Equation (\ref{e.verthydroeq3}))
\begin{eqnarray}
g \Sigmaecinit &=&  \Pec + \frac{1}{8\pi} [\Binit^{2} (\Sigmaecinit) - \Binit^{2} (0)],\label{e.hydroeqinit}\\
g \Sigmaecfin  &=&  \Pec + \frac{1}{8\pi} [\Bfin^{2} (\Sigmaecfin)  - \Bfin^{2} (0)]\label{e.hydroeqfin},
\end{eqnarray}
respectively, where $\Sigmaecinit$ and $\Sigmaecfin$ are the electron capture column depths.  Subtracting Equation (\ref{e.hydroeqfin}) from (\ref{e.hydroeqinit}) and noting that $\Binit (\Sigmaecinit) \gg \Binit (0)$, $\Bfin (\Sigmaecfin)$, and $\Bfin (0)$ by assumptions (2) and (3) gives the total column of matter available as nuclear fuel, 
\begin{equation}\label{e.deltaSigma}
\Sigmaecinit - \Sigmaecfin \approx \frac{\Binit^{2}}{8 \pi g}.
\end{equation}
See Figure \ref{f.schematic} for a schematic diagram.  

Assuming spherical symmetry, the total energy released via electron captures 
\begin{equation}\label{e.energyrelease1}
\Etot \approx 4 \pi R^{2} (\Sigmaecinit - \Sigmaecfin) \frac{Q}{\mb},
\end{equation}
where $Q$ is the energy per nucleon released via electron capture and $\mb$ is the mean nucleon mass.  In general, a nucleus will undergo several electron captures in the outer crust as the field decays, but for most reactions $Q \lesssim 0.1 \nsp \MeV$ \citep{GBSMK07,HZ08}.  Captures that release the most heat occur near neutron drip (where $P \approx 1 \ee{30} \nsp \dyne \usp \power{\cm}{-2}$), i.e., those that induce neutron emissions; \citet{GKM08} find $Q \sim 1 \nsp \MeV$.  From Equations (\ref{e.deltaSigma}) and (\ref{e.energyrelease1}), and normalizing to conditions at neutron drip,                                                                                                                                                                                                                                                                                                                                                                                                                                                                                
\begin{equation}\label{e.energyrelease2}
\Etot \approx 3\ee{47}  \left ( \frac{\Binit}{10^{16} \nsp \gauss} \right )^{2} \left ( \frac{Q}{1 \nsp \MeV} \right ) \left ( \frac{R}{10 \nsp \km} \right )^{4} \left ( \frac{M}{1.4 \nsp \Msun} \right )^{-1}  \nsp \ergs.
\end{equation}

\section{Comparison to Joule Heating}\label{s.jouleheatingcomparison}

The usual explanation for magnetars' persistent thermal emission is Joule heating in the crust \citep[e.g.,][]{MUK98,APM08,UK08,PMG09}:   The crust's electrical resistivity is nonzero, so currents generating the internal magnetic field decay in time, converting some of the magnetic energy to thermal energy and thereby heating the crust.  In this section, we compare heating via field-decay electron captures to Joule heating.

From Equation (\ref{e.deltaSigma}) and the expression for the capture pressure scale height for relativistic, degenerate electrons $\hec = \Pec/ \rho g  = \EFec / 4 \mue \mb g$, where $\mue$ is the mean molecular weight per electron, Equation (\ref{e.energyrelease1}) becomes
\begin{equation}\label{e.E1}
\Etot \approx 4 \pi R^{2} \hec \left ( \frac{\Binit^{2}}{8 \pi} \right ) \left ( \frac{4 \mue Q}{\EFec} \right ).
\end{equation}
The term $4 \pi R^{2} \hec ( \Binit^{2}/8 \pi) \approx \Emag$, the total magnetic energy contained in a crust of scale height $\hec$; the ratio of the total nuclear energy released via field-decay electron captures to the total magnetic energy
\begin{equation}\label{e.energyratio}
\frac{\Etot}{\Emag} \approx  \frac{4 \mue Q}{\EFec} = 0.5  \left ( \frac{\mue}{3.2} \right ) \left ( \frac{Q}{1 \nsp \MeV} \right )  \left ( \frac{\EFec}{27 \nsp \MeV} \right )^{-1}.
\end{equation}
Nuclear heating and Joule heating are comparable if all of the magnetic energy converts to thermal energy.  However, unlike electron captures, the crustal magnetic field powers phenomena other than the thermal emission, such as the hard ($\sim 100 \nsp \keV$) nonthermal X-ray  emission \citep[e.g.,][]{KHM04,MCLRPS04,Retal04}, X-ray bursts, and giant flares.  The fraction of magnetic energy that contributes to the thermal emission is unknown.  Therefore, we conclude that heating via field-decay electron captures is a considerable, if not the primary, source of the persistent soft X-ray luminosity from magnetars.

However, we stress that equations (\ref{e.energyrelease2}-\ref{e.energyratio}) are upper limits with respect to our assumed magnetic field configuration.  If the field were instead predominantly radial or force-free, for example, field decay would induce little nuclear heating.

\section{Discussion and Future Work}\label{s.discussion}

In this work, we identified a new, strong heat source in magnetar crusts.  Magnetic pressure partially supports the crust against gravity.  The electron degeneracy pressure increases as the crust loses magnetic pressure support during field decay, and the concomitant increase in $\EF$ induces exothermic electron captures.  The total heat released via field-decay electron captures is of order the crust's total magnetic energy.  Field-decay electron captures are important to neutron stars' magneto-thermal evolution and should be included in future calculations.  At a minimum, field-decay electron captures supplement Joule heating and lessen the interior magnetic field strength required to power magnetar phenomena; however, since only a fraction of the crust's magnetic energy, but essentially all of the nuclear energy, converts to thermal energy, field-decay electron captures may in fact dominate magnetars' persistent soft X-ray emission.

We now outline other potential ramifications of this work:

(1) Although we neglected both the inner crust and outermost envelope for simplicity, field-decay-induced nuclear reactions can occur there as well.  In addition to electron captures in the inner crust, the loss of magnetic pressure support may trigger light-element fusion reactions in the envelope or pycnonuclear reactions in the inner crust, since $\rho$ increases at a fixed $\Sigma$ during field decay.

(2) The crust's electrical resistivity decreases with $\rho$ and increases with $T$ \citep[e.g.,][]{YU80}, so magnetic field decay via Hall-drift-enhanced Ohmic dissipation \citep[e.g.,][]{GR92} is most efficient at low densities and high temperatures.  Magnetic pressure support lowers $\rho$ at a given $\Sigma$ relative to the nonmagnetic case, and field-decay electron captures heat the crust; both effects accelerate Ohmic dissipation.

(3) Although it is the dominant mechanism, field decay is not required to induce electron captures.  For example, rearrangement of the crustal magnetic field via Hall drift, which is nondissipative, can induce captures:  Altering magnetic pressure gradients alone can either raise or lower the electron pressure, and thereby $\EF$, of a matter parcel past an electron capture threshold.  Raising $\EF$ past a capture threshold triggers exothermic electron captures, but lowering $\EF$ past a threshold does nothing, since exothermic captures are effectively irreversible.  This may further enhance the heating rate, and hence the Ohmic dissipation rate, in the crust.

(4) Asymmetries in the internal magnetic field relative to the stellar rotation axis produce distortions in a young, rapidly rotating magnetar that may generate gravitational waves.  A nascent misalignment between the rotation and magnetic axes is secularly unstable and grows with time \citep[e.g.,][]{C02,DSS09}.  If the distorting magnetic field is predominantly nonradial and within the crust, the resulting field evolution causes electron captures, which can generate asymmetric density variations that intensify gravitational wave emission \citep[see][who proposed a similar mechanism for weakly magnetic, accreting neutron stars]{B98,UCB00}.  

(5) We neglected the crust's elasticity in the hydrostatic equilibrium equation (\ref{e.hydroeq}).  Elasticity partially compensates local magnetic stresses and thereby delays electron captures; conversely, sudden pressure increases that offset fracture-induced elasticity losses during catastrophic events like SGR bursts and giant flares may trigger electron captures.   While unlikely to contribute to the burst's emission itself, captures triggered by crust fracture may contribute to the subsequent transient cooling \citep[e.g.,][]{LET02}.

\acknowledgments 
We thank Jonathan Arons, Lars Bildsten, Edward Brown, Andreas Reisenegger, Matthew van Adelsberg, and Ellen Zweibel for helpful discussions, and the referee for constructive questions and comments.  R.L.C.\ and D.L.K.\ are supported by the National Science Foundation under Grant No.\ NSF PHY05-51164.  D.L.K.\ is supported also by NASA through Hubble Fellowship grant \#01207.01-A awarded by the Space Telescope Science Institute, which is operated by the Association of Universities for Research in Astronomy, Inc., for NASA, under contract NAS 5-26555.

\begin{appendix}
\section{Electron Captures Via Landau Quantization Loss}

\begin{figure}
\epsscale{1.1}
\plotone{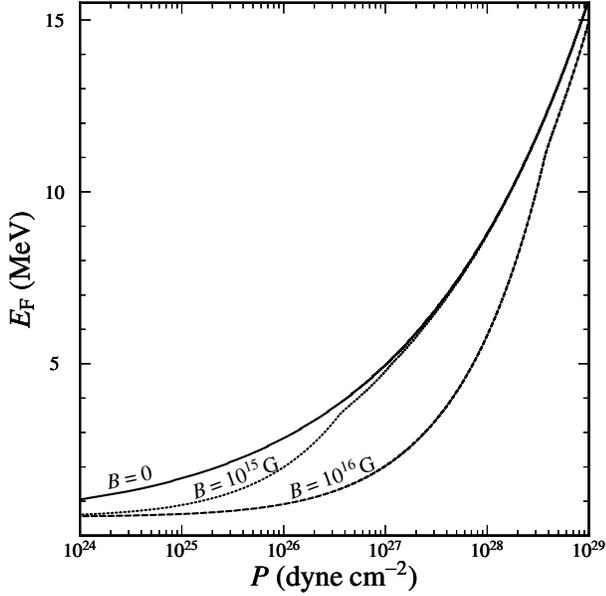}
\caption{Electron Fermi energy $\EF$ as a function of pressure $P = \Pe + \Pc$, where $\Pe$ is the electron degeneracy pressure and $\Pc$ is the Coulomb correction, for different magnetic field strengths $B$.  A magnetic field decreases $\EF$ at fixed $P$ relative to the nonmagnetic case; $\EF$ increases as $B$ decays, inducing electron captures.  The kink in the $B > 0$ plots indicates the $\EF$ value below which electrons occupy only the ground Landau level (Equation \ref{e.EFmax}).  
}
\label{f.EFvsPe}
\end{figure}

A magnetic field quantizes electron motion perpendicular to $\vec{B}$ into Landau orbitals and thereby lowers the electron degeneracy \citep[e.g.,][]{HL06}.  The electron degeneracy pressure $\Pe \propto \nel \EF$, where $\nel$ is the electron number density; for a given $\Pe$, $\vec{B}$ increases $\nel$ and lowers $\EF$ relative to the nonmagnetic case.  Therefore,  $\EF$ increases as the magnetic field decays, inducing electron captures.  The Coulomb correction to the pressure $\Pc \propto - \nel^{4/3}$ negligibly mitigates this effect.

Landau quantization affects $\EF$ only when $P \ll B^{2}/8\pi$ (Figure \ref{f.EFvsPe}).  The electron energy eigenstates 
\begin{equation}\label{e.eigenstates}
E_{\nu, \pperp} = \left [ \pperp^{2} c^{2} + \me^{2} c^{4} \left ( 1 + 2 \nu \frac{B}{\Bcrit} \right )\right ]^{1/2},
\end{equation}
where $\pperp$ is the electron momentum perpendicular to $\vec{B}$, $\nu$ is the quantum number, $\Bcrit = \me^{2} c^{3}/ \hbar e = 4.414\ee{13} \nsp \gauss$ is the quantum critical field strength, and $\me$, $c$, $\hbar$, and $e$ are the usual fundamental constants.  The magnetic field has the largest effect on $P$ when all electrons are in the ground Landau level $\nu = 0$.  From Equation (\ref{e.eigenstates}), this occurs when 
\begin{equation}\label{e.EFmax}
\EF < \me c^{2} \left ( 1 + 2 \frac{B}{\Bcrit} \right )^{1/2} \approx 10.9 \left ( \frac{B}{10^{16} \nsp \gauss} \right )^{1/2} \nsp \MeV.
\end{equation}
From Equation (\ref{e.EFmax}) and the relation $\Pe \approx \EF^{4}/12 \pi^{2} (\hbar c)^{3}$ for relativistic, degenerate electrons at $B=0$ \citep[e.g.,][]{ST83}, Landau quantization affects $P$ when 
\begin{equation}\label{e.PoverPmag}
\frac{\Pe}{B^{2}/8 \pi} \lesssim \frac{8 \alpha}{3\pi} =6 \ee{-3},
\end{equation}
where $\alpha = e^{2}/\hbar c$ is the fine structure constant.   

Unlike captures via magnetic pressure support loss (\S \ref{s.fielddecayecaptures}), captures via Landau quantization loss always occur during magnetic field decay, i.e., they occur independent of the field's geometry.  However, Equation (\ref{e.PoverPmag}) shows that Landau quantization's effect on electron captures is small relative to that of magnetic pressure support if $\vec{B}$ is primarily nonradial.  

\end{appendix}

\bibliographystyle{apj}

\end{document}